\documentclass{amsart}

\usepackage{setspace}

\usepackage[dvips]{epsfig}

\usepackage{amsmath,amssymb,latexsym, amscd}
\usepackage{exscale, cite, eps fig, graphics}

\usepackage[colorlinks]{hyperref}

\renewcommand{\leq}{\leqslant}

\newcommand{\be}{\begin{equation}}  
\newcommand{\ee}{\end{equation}}

\title[Rotational and capillary waves]
{A new application of Crapper's exact solution \\ to waves in constant vorticity flows}

\author[Hur]{Vera~Mikyoung~Hur}
\address{Department of Mathematics, University of Illinois at Urbana-Champaign, Urbana, IL 61801 USA}
\email{verahur@math.uiuc.edu}

\author[Vanden-Broeck]{Jean-Marc~Vanden-Broeck}
\address{Department of mathematics, University of College London, London, WC1E 6BT, UK}
\email{j.vanden-broeck@ucl.ac.uk}

\date{\today}

\begin{document}

\maketitle

\begin{abstract}
In 1957 Crapper found an exact solution for capillary waves propagating at the surface of an irrotational flow of infinite depth. Here we provide conclusive analytical and numerical evidence that a Crapper wave makes the profile of a periodic traveling wave propagating, in the absence of the effects of gravity and surface tension, in a constant vorticity flow. This is achieved by constructing a Stokes expansion up to the third order in a small amplitude parameter and  by numerically computing large amplitude waves.
\end{abstract}

\section{Introduction}\label{sec:intro}

We are interested in periodic traveling waves in a two-dimensional, infinitely deep and constant vorticity flow, and their links to capillary waves. 

For gravity waves in an irrotational flow (zero vorticity), Stokes \cite{Stokes1847, Stokes1880} observed that crests become sharper and troughs flatter as the amplitude increases and that the so-called wave of greatest height or {\em extreme} wave possesses a $120^\circ$ angle at the crest. 
For strongly positive vorticity, by contrast, the numerical computations \cite{SS1985} indicate that the amplitude increases, decreases and increases in the (wave speed) $\times$ (amplitude) plane. Namely, a fold develops. Moreover, the wave profile becomes vertical and overturns as the amplitude increases along the fold, and a limiting configuration is either an {\em extreme} wave, like in an irrotational flow, or a {\em touching} wave, which encloses a bubble of air at the trough. Here we distinguish positive vorticity for waves propagating upstream, whereas negative vorticity for downstream. The latter author \cite{VB1996} numerically found, among others, a new solution branch limited by a touching wave as the gravitational acceleration vanishes, whose crest-to-trough height for the unit period -- namely, the steepness -- is approximately $0.7$.

For capillary waves (nonzero surface tension and zero gravity) in an irrotational flow, on the other hand, Crapper \cite{Crapper1957} produced an exact solution formula for all amplitudes and thereby deduced that crests become flatter and troughs more curved as the amplitude increases toward a touching wave whose steepness is approximately $0.7$. 

This fortuitous coincidence is no accident. 
Recently, Dyachenko and the former author \cite{DH2019a,DH2019b} (see also \cite{DH2019c}) numerically found, among others, that touching waves in positive vorticity flows approach the limiting Crapper wave as the vorticity strength increases indefinitely or, equivalently, the gravitational acceleration vanishes -- an unexpected and remarkable link between rotational and capillary effects. 

Here we offer conclusive analytical and numerical evidence that {\em any} Crapper wave, not necessarily the limiting one, makes the profile of a periodic traveling wave for positive constant vorticity and zero gravity. For weakly nonlinear waves, we verify that a small amplitude expansion of a periodic traveling wave in a constant vorticity flow agrees with that of a capillary wave in an irrotational flow up to the third order, although the flows underneath the fluid surface are different. For strongly nonlinear waves, following \cite{VB1996}, we numerically compute the solutions and compare with Crapper's exact solution \cite{Crapper1957}. It is desirable to offer analytical evidence for all amplitudes. This is an interesting open problem. 

\section{Formulation}\label{sec:formulation}

We consider a two-dimensional, infinitely deep and constant vorticity flow of an incompressible inviscid fluid, or else an irrotational flow under the influence of surface tension, and waves propagating at the fluid surface. We neglect the effects of gravitational acceleration. 
Although an incompressible fluid may have variable density, we assume for simplicity the unit density. 

Suppose for definiteness that in Cartesian coordinates, the $x$ axis points in the direction of wave propagation and the $y$ axis vertically upwards. Suppose that the fluid at time $t$ occupies a region in the $(x,y)$ plane, bounded above by a free surface $y=\eta(x,t)$. We assume for now that $\eta$ is single-valued (but see Section~\ref{sec:numerical}). Let $\mathbf{u}=\mathbf{u}(x,y,t)$ denote the velocity of the fluid at the point $(x,y)$ and time $t$, and $P=P(x,y,t)$ the pressure. They satisfy the Euler equations for an incompressible fluid:
\begin{subequations}\label{E:ww}
\begin{equation}\label{E:euler}
\left.\begin{aligned}
&\mathbf{u}_t+(\mathbf{u}\cdot\nabla)\mathbf{u}=-\nabla P\\
&\nabla\cdot\mathbf{u}=0
\end{aligned}
\right\}\quad\text{in $-\infty<y<\eta(x,t)$}.
\end{equation}
Let
\begin{equation}\label{E:vorticity}
\Omega=\nabla\times\mathbf{u}
\end{equation}
denote constant vorticity. Note that if the vorticity is constant in the fluid region at the initial time then it remains so at all later times. The kinematic and dynamic conditions:
\begin{equation}\label{E:boundary}
\left.\begin{aligned}
&\eta_t+\mathbf{u}\cdot\nabla(\eta-y)=0\\ 
&P=-T\frac{\eta_{xx}}{\sqrt{1+\eta_x^2}^3}
\end{aligned}
\right\}\quad\text{at $y=\eta(x,t)$}
\end{equation}
\end{subequations}
express that each fluid particle at the surface remains so for all time, and that there is a jump in the pressure across the fluid surface, proportional to the curvature, where $T$ is the coefficient of surface tension. 

When $T=0$ (zero surface tension), for any $\Omega$, note that
\begin{equation}\label{E:shear}
\eta(x,t)=0, \quad \mathbf{u}(x,y,t)=(-\Omega y,0)\quad\text{and}\quad P(x,y,t)=0
\end{equation}
solve \eqref{E:ww} for all time. They make a linear shear flow, for which the fluid surface is horizontal and the fluid velocity varies linearly with depth. We assume that some external effects such as wind produce a flow of the kind and restrict the attention to waves propagating in \eqref{E:shear}. 

Suppose that 
\begin{equation}\label{def:velocity}
\mathbf{u}=(-\Omega y,0)+\nabla\varPhi,
\end{equation}
whence the latter equation of \eqref{E:euler} implies that $\nabla^2\varPhi=0$ in $-\infty<y<\eta(x,t)$. 
We pause to remark that for non-constant vorticity, such a velocity potential is no longer viable to use. 
Substituting \eqref{def:velocity} into the former equation of \eqref{E:euler} and using the latter equation of \eqref{E:boundary}, we make an explicit calculation to show that 
\[
\varPhi_t+\frac12(\varPhi_x-\Omega y)^2-\frac12\Omega^2y^2+\frac12\varPhi_y^2+\Omega\Psi
-T\frac{\eta_{xx}}{\sqrt{1+\eta_x^2}^3}=B(t)\quad\text{at $y=\eta(x,t)$}
\]
for some function $B(t)$, where $\varPsi$ be a harmonic conjugate of $\varPhi$. 

We turn the attention to traveling waves of \eqref{E:ww}. That is, $\mathbf{u}$ and $P$ are functions of $(x-ct,y)$ while $\eta$ is a function of $x-ct$ for some $c>0$, the wave speed. Under the assumption, we will go to a moving coordinate frame, changing $x-ct$ to $x$, whereby $t$ completely disappears. The result becomes:
\begin{subequations}\label{E:steady}
\begin{align}
&\nabla^2\varPhi=0 & &\text{in $y<\eta(x)$},\label{E:phi} \\
&(\varPhi_x-\Omega y-c)\eta'=\varPhi_y & &\text{at $y=\eta(x)$}, \label{E:kinematic}\\
&\frac12(\varPhi_x-\Omega y-c)^2+\frac12\varPhi_y^2-T\frac{\eta''}{\sqrt{1+(\eta')^2}^3}=B 
& &\text{at $y=\eta(x)$}\label{E:dynamic}
\intertext{for some constant $B$, where the prime means differentiation in the $x$ variable. The boundary condition in the infinite depth is that}
&\varPhi\to0\quad &&\text{as $y\to-\infty$}.\label{E:infty}
\end{align}
\end{subequations}
When $\Omega=0$ (zero vorticity), \eqref{E:steady} makes the capillary wave problem \cite{Crapper1957}.

In what follows, we assume that $\varPhi$ and $\eta$ are periodic, $\varPhi$ is odd and $\eta$ even  in the $x$ variable. 

\section{Small amplitude: Stokes expansion}\label{sec:expansion}

We proceed to the small amplitude expansion of a solution of \eqref{E:steady} when either $T=0$ (nonzero vorticity and zero surface tension) or $\Omega=0$ (zero vorticity and nonzero surface tension). We may assume without loss of generality that $\eta$ is of mean zero, and let
\begin{equation}\label{def:eta}
\eta(x)=\epsilon \cos(kx)+\epsilon^2\eta_2\cos(2kx)+\epsilon^3\eta_3\cos(3kx)+\cdots,
\end{equation}
where $\epsilon\ll1$ is an amplitude parameter and $k$ the wave number, $\eta_2, \eta_3,\dots$ are to be determined. Let
\begin{equation}\label{def:phi}
\varPhi(x,y)=\epsilon\varPhi_1e^{ky}\sin(kx)+\epsilon^2\varPhi_2e^{2ky}\sin(2kx)
+\epsilon^3\varPhi_3e^{3ky}\sin(3kx)+\cdots
\end{equation}
for $\epsilon\ll1$. Note that \eqref{def:phi} solves \eqref{E:phi} and \eqref{E:infty} for appropriate $\varPhi_1,\varPhi_2,\varPhi_3, \dots$. Moreover, let
\begin{equation}\label{def:c}
c=c_0+\epsilon^2c_2+\cdots
\end{equation}
for $\epsilon\ll1$, where $c_0, c_2, \dots$ are to be determined.
We will use \eqref{E:kinematic}-\eqref{E:dynamic} to determine \eqref{def:eta}-\eqref{def:c} up to the order of $\epsilon^3$ when either $T=0$ or $\Omega=0$. In the latter, one may resort to Crapper's exact solution~\cite{Crapper1957} (see \eqref{E:Crapper}), but it is simpler and more instructive to directly use the equations. 

Substituting \eqref{def:eta}-\eqref{def:c} to \eqref{E:kinematic}-\eqref{E:dynamic}, at the order of $\epsilon$, we find that 
\begin{equation}\label{E:phi1}
\varPhi_1=c_0=\begin{cases}
\Omega/k\quad &\text{for $T=0$},\\
\sqrt{Tk} &\text{for $\Omega=0$}.
\end{cases}
\end{equation}
The result agrees with \cite[(3.4)]{SS1985} in the absence of the gravitational acceleration. It agrees with \cite[(3.7)]{HFMK2016} (see also references therein) for gravity-capillary waves in a constant vorticity flow of finite depth, as the fluid depth increases indefinitely, after returning to dimensional variables.

At the order of $\epsilon^2$, \eqref{E:kinematic}-\eqref{E:dynamic} become
\begin{align*}
c_0\eta_2-\varPhi_2=&\frac14c_0k-\frac14(c_0k-\Omega),\\
c_0(2k\varPhi_2-\Omega\eta_2)-4k^2T\eta_2=&-\frac34c_0^2k^2+\frac14(c_0k-\Omega)^2.
\end{align*}
Here we use \eqref{def:phi} and \eqref{def:eta}, and we make a Taylor series expansion to find that
\begin{align*}
\varPhi_x(x,\eta(x))=&\epsilon k\varPhi_1\cos(kx)+\epsilon^2k^2\varPhi_1\cos^2(kx)+2\epsilon^2k\varPhi_2\cos(2kx)\\
&+\epsilon^3k^2\varPhi_1\eta_2\cos(kx)\cos(2kx)+\tfrac12\epsilon^3k^3\varPhi_1\cos^3(kx)\\
&+4\epsilon^3k^2\varPhi_2\cos(kx)\cos(2kx)+3\epsilon^3k\varPhi_3\cos(3kx)+\cdots,
\end{align*}
and similarly for $\varPhi_y(x,\eta(x))$. We use \eqref{E:phi1} and make an explicit calculation to find that 
\begin{equation}\label{E:eta2}
\eta_2=-\frac14k\quad\text{when either $T=0$ or $\Omega=0$},
\end{equation}
and 
\begin{equation}\label{E:phi2}
\varPhi_2=\begin{cases}
-\frac12c_0k\quad&\text{for $T=0$}\\
-\frac34c_0k&\text{for $\Omega=0$}.
\end{cases}
\end{equation}
Therefore, the profile of a periodic traveling wave in a constant vorticity flow coincides with that of a capillary wave in an irrotational flow up to the second order of a small amplitude parameter, although the flows underneath the fluid surface are different. The result agrees with \cite[(3.17)-(3.18)]{HFMK2016} in the infinite depth limit, after adaptation to the present notation.

To continue, at the order of $\epsilon^3$, \eqref{E:kinematic}-\eqref{E:dynamic} become
\begin{align*}
3c_0k\eta_3-3k\varPhi_3=&\frac14c_0k^2-\frac14k(c_0k-\Omega)+2k\varPhi_2+\frac12(2k\varPhi_2-\Omega\eta_2),\\
c_0(3k\varPhi_3-\Omega\eta_3)-9k^2T\eta_3=&-\frac14c_0^2k^3+\frac14c_0k^2(c_0k-\Omega)\\
&-3c_0k^2\varPhi_2+\frac12(c_0k-\Omega)(2k\varPhi_2-\Omega\eta_2)+\frac98Tk^4.
\end{align*}
We use \eqref{E:phi1}, \eqref{E:eta2}, \eqref{E:phi2} and make an explicit calculation to find that
\begin{equation}\label{E:eta3}
\eta_3=\frac{1}{16}k^2\quad\text{when either $T=0$ or $\Omega=0$},
\end{equation}
while
\[
\varPhi_3=\begin{cases}
\frac{7}{16}c_0k^2\quad&\text{for $T=0$}\\
\frac{13}{16}c_0k^2&\text{for $\Omega=0$}.
\end{cases}
\]
Therefore, the profiles of a periodic traveling wave in a constant vorticity flow and a capillary wave in an irrotational flow coincide up to the third order of a small amplitude parameter. The result agrees\footnote{The steepness is used as a small amplitude parameter in \cite{HFMK2016} whereas here we use the first Fourier coefficient, notwithstanding different notation.} with \cite[(3,28) and (3.30)]{HFMK2016} in the infinite depth limit.

\section{Arbitrary amplitude: numerical computation}\label{sec:numerical}

We take matters of weakly nonlinear theory in Section \ref{sec:expansion} to strongly nonlinear. In what follows, let
\[
x\mapsto \frac{k}{2\pi}x\quad\text{and}\quad y\mapsto \frac{k}{2\pi}y,
\]
where $k$ is the wave number, and we use dimensionless variables
\begin{equation}\label{def:no-dimension}
\phi=\frac{k}{2\pi}\frac{\varPhi}{c}\quad\text{and}\quad \omega=\frac{2\pi}{k}\frac{\Omega}{c}.
\end{equation}
We allow that the fluid surface is multi-valued, described in parametric form. We may assume without loss of generality that the crest is at the origin in the $(x,y)$ plane. 

When $\omega=0$ (the capillary wave problem), recall Crapper's exact solution \cite{Crapper1957}, written $x=X(\phi)$ and $y=Y(\phi)$, $0\leq\phi\leq1$, where
\be\label{E:Crapper}
\left\{
\begin{aligned}
&X(\phi)=\phi-\frac{2}{\pi}\frac{A\sin2\pi\phi}{1+A^2-2A\cos2\pi\phi}, \\ 
&Y(\phi)=-\frac{2}{\pi}\frac{1}{1+A}+\frac{2}{\pi}\frac{1+A\cos2\pi\phi}{1+A^2-2A\cos2\pi\phi},
\end{aligned}
\right.
\ee
and
\[
A=\frac{2}{\pi s} \left(\left(1+\frac14\pi^ 2 s^2\right)^{1/2}-1\right).
\]
Here and elsewhere, $s$ denotes the steepness, the crest-to-trough height divided by the period. 

\begin{figure}
	\centering
	\includegraphics[width=8cm,angle=-90]{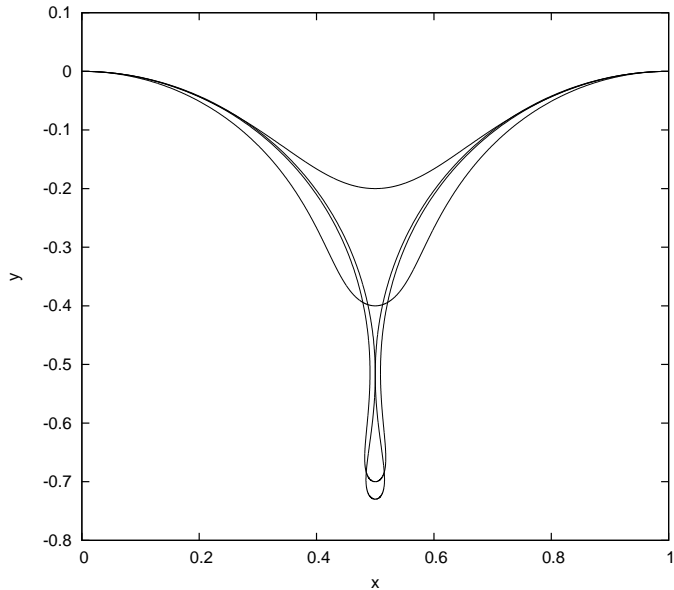}
	\caption{Profiles of Crapper waves, and also periodic traveling waves in constant vorticity flows, for $s=0.2$, 
	$0.4$, $0.7$ and $0.7298$. For constant vorticity, the corresponding values of $\omega$ are $5.47$, $3.91$, $2.14$ and $2.02$.}
	\label{fig} 
	\end{figure}

Figure~\ref{fig} provides wave profiles for several values of $s$. When $s$ is small, the profile is found single-valued and well approximated by \eqref{def:eta}, \eqref{E:eta2} and \eqref{E:eta3}, after returning to dimensionless variables. As $s$ increases, we confirm that crests become more rounded and troughs more curved, and as $s\to s^*\approx0.7298$, the profile approaches a limiting configuration which encloses a bubble of air at the trough. For $s>s^*$, the fluid surface intersects itself and the flow becomes multi-valued, whence the solution is not physically realistic. 

When $T=0$ (the constant vorticity problem), on the other hand, there is no exact solution formula to the best of the authors' knowledge, and we follow \cite{VB1996} (see also \cite{SS1985,VB1995}) to numerically compute the solutions. Here we discuss the main ideas for completeness and for some differences in the choice of the parameters. Readers are referred to \cite{VB1996} and references therein for details. 

Suppose that $x=X(\tau)$ and $y=Y(\tau)$ parametrize the fluid surface, where $\tau$ is the arclength, and that $\tau=0$ at the crest:
\begin{subequations}\label{E:reformulation}
\begin{gather}
X'(\tau)^2+Y'(\tau)^2=1\quad\text{for $-L/2\leq\tau\leq L/2$},\label{E:XY} 
\intertext{where $L$ is the length of the fluid surface over one period, and}
X(0), Y(0)=0. \label{E:crest}
\end{gather}
Let
\[
u=\phi_x \quad {\hbox{and}} \quad v=\phi_y.
\]
Note that $u-iv$ is an analytic function of $x+iy$ and $u-iv\to0$ as $y\to-\infty$, and let 
\[
(u-iv)(\tau)=(u-iv)(X(\tau)+iY(\tau)).
\] 
An application of the Cauchy integral formula leads to 
\begin{equation}\label{E:u}
\begin{aligned}
\pi u(\tau)=&-\text{Im}\int_0^{L/2}
\frac{(u(\sigma)-iv(\sigma))(-2\pi i X'(\sigma)+2 \pi Y'(\sigma))}{1-\exp(-2\pi i(X(\tau)-X(\sigma))+2\pi(Y(\tau)-Y(\sigma)))}~d\sigma \\
&-\text{Im} \int_0^{L/2}
\frac{(u(\sigma)+iv(\sigma))(-2\pi i X'(\sigma)-2\pi Y'(\sigma))}{1-\exp(-2\pi i(X(\tau)+X(\sigma))+2\pi(Y(\tau)-Y(\sigma)))}~d\sigma.
\end{aligned}
\end{equation}
See \cite{VB1996} for details. Moreover, \eqref{E:kinematic} and \eqref{E:dynamic} become
\begin{gather}
(u-1-\omega Y)Y'(\tau)=vX'(\tau), \label{E:kinematic'} \\
(u-1-\omega Y)^2(\tau)+v^2(\tau)=B \label{E:dynamic'}
\end{gather}
\end{subequations}
for $-L/2\leq\tau\leq L/2$.

\subsection{Numerical procedure}\label{sec:method}

For $\omega$ given, we seek a numerical solution, $u$ and $v$, $X'$ and $Y'$, of \eqref{E:reformulation}, where $B$ and $L$ are determined as part of the solution.

Let
\[
\tau_j=\frac{L}{2}\frac{j-1}{N-1},\quad j=1,2,\dots,N,
\]
define $N$ uniform mesh points over $[0,L/2]$, and let
\begin{subequations}\label{def:unknowns}
\begin{gather} 
u_j=u(\tau_j)\quad\text{and}\quad v_j=v(\tau_j),\quad j=1,2,\dots,N, \label{def:uvj}\\
X'_j=X'(\tau_j)\quad\text{and}\quad Y'_j=Y'(\tau_j),\quad j=1,2,\dots,N,\label{def:XY'j}
\end{gather}
\end{subequations}
make $4N$ unknowns. Let 
\[
\tau_{j-1/2}=\frac{\tau_{j+1}+\tau_j}{2}, \quad j=1,2,\dots,N-1.
\]
We approximate $X_j=X(\tau_j)$ and $Y_j=Y(\tau_j)$, $j=1,2,\dots, N$, using \eqref{def:XY'j} and the trapezoidal rule. That is, $X_1=0$, $Y_1=0$ (see \eqref{E:crest}), and
\begin{align*}
X_j&=X_{j-1}+X'(\tau_{j-3/2})\frac{L}{2}\frac{1}{N-1},  \\
Y_j&=Y_{j-1}+Y'(\tau_{j-3/2})\frac{L}{2}\frac{1}{N-1},
\end{align*}
$j=2,3,\dots,N$, where we evaluate $X'(\tau_{j-3/2})$ and $Y'(\tau_{j-3/2})$ using \eqref{def:XY'j} and 
a four-point interpolation formula. 

We satisfy \eqref{E:XY}, \eqref{E:kinematic'} and \eqref{E:dynamic'} at $\tau=\tau_j$, $j=1,2,\dots,N$. They make $3N$ equations. We then satisfy \eqref{E:u} at $\tau=\tau_j$, $j=2,3,\dots,N-1$, using the trapezoidal rule and summing over $\sigma=\tau_{j-1/2}$, $j=1,2,\dots,N-1$. This makes $N-2$ equations. We pause to remark that the symmetry of the discretization and of the trapezoidal rule with respect to the singularity of the integrand at $\sigma=\tau$ enables us to evaluate the Cauchy principal value integral with an accuracy no less than a non-singular integral. Lastly, we require two symmetry conditions
\[
v_1, v_N=0.
\]
In addition, since the crest is at the origin, 
\[
Y_N=-s,
\] 
where $s$ is the steepness. Since $u-iv$ vanishes in the infinite depth, 
\[
\int_0^{L/2} (uX'+vY')(\tau)~d\tau=0.
\]
Moreover, we may require 
\[
X_N=1/2
\] 
-- namely, the unit period. Together, they make $4N+3$ equations for $4N+3$ unknowns \eqref{def:unknowns}, plus $B$, $L$ and $\omega$. We will solve them by Newton's method, with $s$ a parameter. See \cite{VB1996} for details.

\subsection{Numerical results}\label{sec:results}

We followed Section~\ref{sec:method} and numerically computed the solutions of \eqref{E:reformulation} for various values of $s$ and $N$. For $N>200$, wave profiles were found indistinguishable within graphical accuracy from Crapper waves. See, for instance, Figure~\ref{fig} for $s=0.2$, $0.4$, $0.7$ and $0.7298$. The corresponding values of $\omega$ were found to be $5.47$, $3.91$, $2.14$ and $2.02$. 
As $s\rightarrow 0$, we found that $\omega \rightarrow 2\pi$, whence $\Omega\to ck$ (see the latter equation of \eqref{def:no-dimension}). The result agrees with \eqref{E:phi1}. 

To make comparisons more quantitative, we computed the length of a Crapper wave over one period, by numerically evaluating 
\begin{equation}\label{E:LCrapper}
L=2\int_0^{1/2} \sqrt{X'(\phi)^2+Y'(\phi)^2}~d\phi,
\end{equation}
where $X$ and $Y$ are in \eqref{E:Crapper}, 
using the trapezoidal rule, which is spectrally accurate for periodic functions. Table~\ref{table} provides a comparison of the lengths of the fluid surfaces of the numerical solutions of \eqref{E:reformulation} and the numerical evaluation of \eqref{E:LCrapper}.

\begin{table}
\begin{center}
\begin{tabular}{ l l l l }
\hline
 & \hspace*{5pt} $s=0.2$ & \hspace*{5pt} $s=0.4$ & \hspace*{5pt} $s=0.7$ \\ \hline
$N=100$ & $1.09640843$ & $1.36221253$ & $1.97561843$ \\
$N=200$ & $1.09638286$ & $1.36206379$ & $1.97324907$ \\
$N=300$ & $1.09637804$ & $1.36203852$ & $1.97282934$ \\
$N=400$ & $1.09637633$ & $1.36203002$ & $1.97269855$ \\
$N=500$ & $1.09637553$ & $1.36202619$ & $1.97264290$ \\ 
Crapper &  $1.09637405$ & $1.36201963$ & $1.97255882$ \\
$(N=\infty)$ \\
\hline
\end{tabular}
\caption{Lengths for the numerical solutions in constant vorticity flows vs. Crapper's exact solution.}\label{table}
\end{center}
\end{table}

\subsection*{Acknowledgement}
VMH is supported by the US National Science Foundation under the Faculty Early Career Development (CAREER) Award DMS-1352597. The research was partly done while she visited the University College London (UCL) as a Simons Visiting Professor. She thanks the Simons Foundation and the Mathematisches Forschungsinstitut Oberwolfach for their financial support and the UCL Department of Mathematics for its hospitality. 

\bibliographystyle{amsplain}
\bibliography{vorticity}

\providecommand{\bysame}{\leavevmode\hbox to3em{\hrulefill}\thinspace}
\providecommand{\MR}{\relax\ifhmode\unskip\space\fi MR }
\providecommand{\MRhref}[2]{%
  \href{http://www.ams.org/mathscinet-getitem?mr=#1}{#2}
}
\providecommand{\href}[2]{#2}
\begin{thebibliography}{10}

\bibitem{Crapper1957}
G.~D. Crapper, \emph{An exact solution for progressive capillary waves of
  arbitrary amplitude}, J. Fluid Mech. \textbf{2} (1957), 532--540.
  \MR{0091075}

\bibitem{DH2019c}
Sergey~A. Dyachenko and Vera~Mikyoung Hur, \emph{Stokes waves in a constant
  vorticity flow}, Nonlinear Water Waves, Tutorials, Schools, and Workshops in
  the Mathematical Sciences, Birkh\"auser, Cham, 2019, pp.~71--86.

\bibitem{DH2019b}
\bysame, \emph{Stokes waves with constant vorticity: folds, gaps and fluid
  bubbles}, J. Fluid Mech. \textbf{878} (2019), 502--521. \MR{4010456}

\bibitem{DH2019a}
\bysame, \emph{Stokes waves with constant vorticity: {I}. {N}umerical
  computation}, Stud. Appl. Math. \textbf{142} (2019), no.~2, 162--189.
  \MR{3915685}

\bibitem{HFMK2016}
Hung-Chu Hsu, Marc Francius, Pablo Montalvo, and Christian Kharif,
  \emph{Gravity-capillary waves in finite depth on flows of constant
  vorticity}, Proc. A. \textbf{472} (2016), no.~2195, 20160363, 19.
  \MR{3592261}

\bibitem{SS1985}
J.~A. Simmen and P.~G. Saffman, \emph{Steady deep-water waves on a linear shear
  current}, Stud. Appl. Math. \textbf{73} (1985), no.~1, 35--57. \MR{797557}

\bibitem{Stokes1847}
George~Gabriel Stokes, \emph{On the theory of oscillatory waves}, Trans. Camb.
  Philos. Soc. \textbf{8} (1847), 441--455.

\bibitem{Stokes1880}
\bysame, \emph{Considerations relative to the greatest height of oscillatory
  irrotational waves which can be propagated without change of form},
  Mathematical and Physical Papers, vol.~1, Cambridge University Press, 1880,
  pp.~314--326.

\bibitem{VB1995}
J.-M. Vanden-Broeck, \emph{New families of steep solitary waves in water of
  finite depth with constant vorticity}, European J. Mech. B Fluids \textbf{14}
  (1995), no.~6, 761--774. \MR{1364730}

\bibitem{VB1996}
\bysame, \emph{Periodic waves with constant vorticity in water of infinite
  depth}, IMA J. Appl. Math. \textbf{56} (1996), no.~2, 207--217.

\end{thebibliography}

\end{document}